\documentclass[twocolumn,prb,superscriptaddress]{revtex4}

\usepackage{graphicx}% Include figure files
\usepackage{dcolumn}% Align table columns on decimal point
\usepackage{bm}% bold math
\usepackage{amssymb, amsmath}
\usepackage{color}
\usepackage{slashed} 

\begin{document}
\title{Thermodynamic properties of the Dynes superconductors}

\author{Franti\v{s}ek Herman} 

\affiliation{Department of Experimental Physics, Comenius University,
  Mlynsk\'{a} Dolina F2, 842 48 Bratislava, Slovakia}

\affiliation{Institute for Theoretical Physics, ETH Zurich, CH-8093
  Zurich, Switzerland}

\author{Richard Hlubina}
\affiliation{Department of Experimental Physics, Comenius University,
  Mlynsk\'{a} Dolina F2, 842 48 Bratislava, Slovakia}

\begin{abstract}
The tunneling density of states in dirty superconductors is often well
described by the phenomenological Dynes formula. Recently we have
shown that this formula can be derived, within the coherent potential
approximation, for superconductors with simultaneously present
pair-conserving and pair-breaking impurity scattering.  Here we
demonstrate that the theory of such so-called Dynes superconductors is
thermodynamically consistent. We calculate the specific heat and
critical field of the Dynes superconductors, and we show that their
gap parameter, specific heat, critical field, and penetration depth
exhibit power-law scaling with temperature in the low-temperature
limit. We also show that, in the vicinity of a coupling
constant-controlled superconductor to normal metal transition, the
Homes law is replaced by a different, pair-breaking dominated scaling
law.
\end{abstract}
\maketitle

\section{Introduction}
It is well known that, in the limit of low temperatures, the tunneling
density of states of dirty superconductors $N(\omega)$ is often well
described by the phenomenological Dynes
formula:\cite{Dynes78,Noat13,Szabo16}
\begin{equation}
N(\omega)=N_0{\rm Re}\left[\frac{\omega+i\Gamma}
{\sqrt{(\omega+i\Gamma)^2-{\Delta}^2}}\right],
\label{eq:dynes}
\end{equation}
where $N_0$ is the normal-state density of states, $\Delta$ is the
superconducting order parameter, and $\Gamma$ measures the number of
in-gap states.  The square root in Eq.~(\ref{eq:dynes}) has to be
taken so that its imaginary part is positive and we keep this
convention throughout this paper.

Recently we have shown,\cite{Herman16} within the coherent potential
approximation (CPA), that Eq.~(\ref{eq:dynes}) applies to
superconductors in which, in addition to the usual pair-conserving
disorder, additional classical pair-breaking disorder field is also
present. The crucial assumption which we had to make was that the
pair-breaking potentials were described by a Lorentzian distribution
with width $\Gamma$.  The resulting $2\times 2$ Nambu-Gor'kov electron
Green's function is described by three energy scales: the order
parameter $\Delta$, the pair-conserving scattering rate $\Gamma_s$,
and the pair-breaking scattering rate $\Gamma$:\cite{Herman17}
\begin{equation}
\hat{G}({\bf k},\omega)= \frac{1}{2}\slashed{\partial}
\ln \left[\varepsilon_{\bf k}^2-\epsilon(\omega)^2\right],
\label{eq:dynes_green}
\end{equation}
where $\slashed{\partial} = \tau_0\frac{\partial}{\partial\omega}
-\tau_1\frac{\partial}{\partial\Delta}
-\tau_3\frac{\partial}{\partial\varepsilon_{\bf k}^{}}$ resembles the
Feynman slash derivative, $\tau_i$ are the Pauli matrices, and
\begin{equation}
\epsilon(\omega)
=\sqrt{(\omega+i\Gamma)^2-\Delta^2}+i\Gamma_s.
\label{eq:epsilon}
\end{equation}

In Ref.~\onlinecite{Herman17} we have argued that
Eqs.~(\ref{eq:dynes_green},\ref{eq:epsilon}) represent the simplest
internally consistent extension of the BCS theory to superconductors
with simultaneously present pair-conserving and pair-breaking
processes, and we have called such systems the Dynes
superconductors. Single-particle and electromagnetic properties of the
Dynes superconductors have been studied in detail in
Refs.~\onlinecite{Herman17} and~\onlinecite{Herman17b}, respectively.
In those works, the Dynes phenomenology has been found to be
successful in fitting such different experiments as the angle-resolved
photoemission spectroscopy in the nodal region of optimally doped
cuprates,\cite{Kondo15} and the temperature- and frequency-dependent
optical conductivity of thin MoN films.\cite{Simmendinger16}

The goal of this paper is to extend the theory of the Dynes
superconductors by studying their thermodynamic properties such as the
specific heat and thermodynamic critical field. To this end, in
Sec.~II we will start by demonstrating that the CPA equations are
thermodynamically consistent, since they can be derived from a
free-energy functional of the self-energy. Making use of this
functional, we will derive a convenient expression for the
condensation energy of the Dynes superconductors. In Sec.~III we
compare the Dynes and the BCS thermodynamics. We concentrate on the
low-temperature limit and we show that in this limit the gap
parameter, specific heat, critical field, and penetration depth of the
Dynes superconductors exhibit power-law scaling with temperature, even
in case of s-wave pairing. Moreover, close to $T_c$, we derive the
Ginzburg-Landau functional for the Dynes superconductors.

Next we study how the low-temperature value of the superfluid density
$n_s$ of a Dynes superconductor scales with the transition temperature
$T_c$, when the pairing interaction decreases and the superconductor
to normal metal transition is approached.  This calculation is
motivated by the recent experimental study of superfluid density $n_s$
in overdoped cuprates,\cite{Bozovic16} which finds that the universal
Homes law\cite{Dordevic13} $n_s\propto T_c$ breaks down in strongly
overdoped cuprates, and in the immediate vicinity of the critical
doping a different scaling, $n_s\propto T_c^2$, is observed.  In
Sec.~IV we show that, surprisingly, the s-wave Dynes superconductors
exhibit the same phenomenology.

\section{Free-energy functional}
As observed in an important work by Jani\v{s},\cite{Janis89} the CPA
equations of a non-superconducting (normal) system with impurities can
be derived variationally from an appropriately chosen averaged free
energy functional ${\cal F}$. In this Section we will generalize the
approach of Jani\v{s} to the superconducting state.

We consider a single band of electrons interacting via a local
attractive potential ${\cal U}$, which is supposed to generate a
spatially uniform mean-field pairing potential $\Delta$. In addition,
the electrons are supposed to be subject to random spatially
uncorrelated pair-conserving and pair-breaking on-site fields $U$ and
$V$ with distribution functions $P(U)$ and $P_m(V)$,
respectively. When later specializing to the case of the Dynes
superconductors, for $P_m(V)$ we will take a Lorentzian with width
$\Gamma$.  In the Nambu-Gor'kov notation, on the electrons therefore
acts at each lattice site $l$ the local potential
$$
{\hat V}_l=\Delta\tau_1+U_l\tau_3+V_l\tau_0.
$$ 

In CPA we assume that the averaged electron self-energy ${\hat
  \Sigma}_n$ is local, i.e. only frequency-dependent: the index $n$
stands for the Matsubara frequency $\omega_n$.

Let $\hat{G}_0^{-1}(n{\bf k})=i\omega_n\tau_0-\varepsilon_{\bf
  k}\tau_3$ be the Nambu-Gor'kov Green's function of the clean
non-interacting system, which depends on $\omega_n$ and the electron
momentum ${\bf k}$.  The averaged full Green's function then is
$$
\hat{G}^{-1}(n{\bf  k})=\hat{G}_0^{-1}(n{\bf k})-{\hat \Sigma}_n.
$$ 
As usual, the self-energy is assumed to depend on two real functions
of $\omega_n$, namely the wave-function renormalization
$Z_n$ and the gap function $\Delta_n$:
$$
{\hat \Sigma}_n=i\omega_n(1-Z_n)\tau_0+Z_n\Delta_n\tau_1.
$$ 

In addition to ${\hat \Sigma}_n$, following Jani\v{s} we introduce
another independent variable, the averaged local Green's function
$\hat{\cal G}_n^{-1}$, which also depends only on frequency.  Note
that in the superconducting state both ${\hat \Sigma}_n$ and
$\hat{\cal G}_n^{-1}$ depend parametrically on the pairing potential
$\Delta$.

Following Jani\v{s}, we seek a free energy functional
$$
{\cal F}={\cal F}\left[\Delta,{\hat \Sigma}_n(\Delta),\hat{\cal
    G}_n^{-1}(\Delta)\right]
$$ 
with the property that its minimization with respect to ${\hat
  \Sigma}_n$ and $\hat{\cal G}_n^{-1}$ yields the CPA equations. One
checks readily that the following free energy per lattice site has the
required properties:
\begin{eqnarray}
{\cal F}&=&
-\frac{T}{\cal N}\sum_{n{\bf k}} {\rm Tr} \ln \hat{G}^{-1}(n{\bf k})
+T\sum_n {\rm Tr} \ln \hat{\cal G}_n^{-1}
\nonumber
\\
&-&T\sum_n\left\langle {\rm Tr}\ln(\hat{\cal G}_n^{-1}
-\hat{V}+\hat{\Sigma}_n\right\rangle
+\frac{|\Delta|^2}{\cal U},
\label{eq:free}
\end{eqnarray}
where ${\cal N}$ is the number of lattice sites and the angular
brackets denote averaging with respect to $U$ and $V$.

In fact, by taking the functional derivatives with respect to ${\hat
  \Sigma}_n$ and $\hat{\cal G}_n^{-1}$, we obtain
\begin{eqnarray}
\hat{\cal G}_n&=&\frac{1}{\cal N}\sum_{\bf k}\hat{G}(n{\bf k}),
\nonumber
\\
\hat{\cal G}_n&=&\left\langle \left(\hat{\cal G}_n^{-1}
-\hat{V}+\hat{\Sigma}_n\right)^{-1}\right\rangle.
\label{eq:cpa}
\end{eqnarray}
The first of these equations is consistent with the identification of
$\hat{\cal G}_n$ as a local Green's function, whereas the second one
can be shown to be equivalent to the CPA equation~(4) of
Ref.~\onlinecite{Herman16}.

Finally, minimization of ${\cal F}$ with respect to $\Delta$ yields
the gap equation for the Dynes superconductors\cite{note_gap_equation}
\begin{equation}
\Delta=g\pi T \sum_{\omega_n=-\Omega_{\rm max}}^{\Omega_{\rm max}}
\frac{\Delta_n}{\sqrt{\omega_n^2+\Delta_n^2}},
\label{eq:self-consistent}
\end{equation}
where $g=N_0{\cal U}$ is the dimensionless pairing interaction, $N_0$
is the normal-state density of states per lattice site at the Fermi
level, and $\Omega_{\rm max}$ is the cutoff in frequency space.  Note
that Eq.~(\ref{eq:self-consistent}) agrees with Eq.~(D4) of
Ref.~\onlinecite{Herman16}.

Having established the validity of the free energy functional
Eq.~\eqref{eq:free}, our next goal will be to evaluate the free energy
in the normal and superconducting states. Let us start with the normal
state $\Delta=0$. In this case the CPA self-energy reduces
to\cite{Herman16} ${\hat \Sigma}_n=-i{\rm sgn}(\omega_n)\Gamma_{\rm
  tot}\tau_0$ with a frequency-independent total scattering rate
$$
\Gamma_{\rm  tot}=\Gamma+\Gamma_s.
$$ 
If the density of states can be taken as a constant in the vicinity of
the Fermi level, the normal-state free energy ${\cal F}_N$ has to be
equal to its value in the clean system,
\begin{equation}
{\cal F}_N=-N_0T\int d\varepsilon \ln(1+e^{-\varepsilon/T}).
\label{eq:free_normal}
\end{equation}

Next we evaluate the free energy difference between the
superconducting and the normal state, $\delta {\cal F}={\cal
  F}_S-{\cal F}_N$. According to Eq.~\eqref{eq:free}, $\delta {\cal
  F}$ is a sum of four terms. The first term
contributes\cite{note_1stterm}
\begin{eqnarray*}
-T\sum_n N_0\int_{-\infty}^\infty d\varepsilon
\ln\frac{Z_n^2(\omega_n^2+\Delta_n^2)+\varepsilon^2}
{(Z_n^N\omega_n)^2+\varepsilon^2}
=\\
-2N_0\pi T\sum_n\left[Z_n\sqrt{\omega_n^2+\Delta_n^2}-Z_n^N|\omega_n|\right],
\end{eqnarray*}
where $Z_n^N$ is the normal-state limit of the wave-function
renormalization $Z_n$. In Appendix~A we will show that the
contributions of the second and third terms in Eq.~\eqref{eq:free} can
be neglected in case of a Dynes superconductor. Finally, making use of
the self-consistent Eq.~\eqref{eq:self-consistent}, the contribution
of the fourth term in Eq.~\eqref{eq:free} to $\delta {\cal F}$ can be
written as
$$
\frac{\Delta^2}{\cal U}=N_0\pi T\sum_n\frac{\Delta\Delta_n}
{\sqrt{\omega_n^2+\Delta_n^2}}.
$$
Collecting all terms, the free-energy difference $\delta {\cal F}$ can be 
written in a Bardeen-Stephen like form,\cite{Bardeen64}
$$
\delta {\cal F}=-2N_0\pi T\sum_n
\left[Z_n\sqrt{\omega_n^2+\Delta_n^2}-Z_n^N|\omega_n|-\frac{\Delta\Delta_n}
{2\sqrt{\omega_n^2+\Delta_n^2}}\right].
$$

In order to proceed, let us note that the Eliashberg functions $Z_n$
and $\Delta_n$ for the Dynes superconductors are given
by the explicit expressions\cite{Herman16}
\begin{eqnarray}
Z_n&=&\left(1+\frac{\Gamma}{|\omega_n|}\right)
\left(1+\frac{\Gamma_s}{\Omega_n}\right),
\label{eq:dynes_z}
\\
\Delta_n&=&\frac{|\omega_n|}{|\omega_n|+\Gamma}\Delta,
\label{eq:dynes_delta}
\end{eqnarray}
where we have introduced an auxiliary quantity
$$
\Omega_n=\sqrt{(|\omega_n|+\Gamma)^2+\Delta^2}.
$$ 
Making use of Eqs.~(\ref{eq:dynes_z},\ref{eq:dynes_delta}), the
free energy difference $\delta {\cal F}$ of a Dynes superconductor can
be finally written in a particularly simple form
\begin{equation}
\delta {\cal F}=-N_0\pi T\sum_n
\frac{\left[\Omega_n-(|\omega_n|+\Gamma)\right]^2}{\Omega_n}.
\label{eq:free_diff}
\end{equation}
It is worth pointing out that the free energy difference
Eq.~\eqref{eq:free_diff} does not depend on the pair-conserving
scattering rate $\Gamma_s$. This was not obvious a priori, since
$\Gamma_s$ does enter both $Z_n$ and $Z_n^N$. However, the
independence of thermodynamic properties on $\Gamma_s$ is fully in
accord with Anderson's theorem\cite{Anderson59} and it is pleasing to
observe that CPA is consistent with this rather general theorem.

\section{Thermodynamics of the Dynes superconductors}

\subsection{Low-temperature behavior}
The density of states at the Fermi level of a Dynes superconductor is
finite, $N(0)=N_0\Gamma/\sqrt{\Delta^2+\Gamma^2}$. We have shown in
Ref.~\onlinecite{Herman17b} that this feature results, inter alia, in
a finite microwave absorption in the low-frequency limit down to the
lowest temperatures. The goal of this Subsection is to show that
similar low-temperature anomalies are present also in the
thermodynamic properties of the Dynes superconductors.

In what follows, we will make frequent use of the following
low-temperature Sommerfeld-like expansion, valid to order $T^3$:
$$
2\pi T\sum_{\omega_n>0}^{\Omega} F(\omega_n)=
\int_0^{\Omega} d\omega F(\omega)
+\frac{\pi^2}{6}T^2\left[F'(0)-F'(\Omega)\right],
$$
where $F^\prime(\omega)$ denotes the derivative of $F(\omega)$
with respect to $\omega$. The derivation of the Sommerfeld-like expansion
is sketched in Appendix~B.

{\it Superconducting gap}. The order parameter $\Delta(T)$ can be
found from the self-consistent equation~\eqref{eq:self-consistent},
where on the right-hand side Eq.~\eqref{eq:dynes_delta} is used. In
Ref.~\onlinecite{Herman16} we have shown that the $T=0$ value of the
gap $\Delta(0)$, in a system with pair-breaking scattering rate
$\Gamma$, is given by
$$
\Delta(0)=\sqrt{\Delta_{00}(\Delta_{00}-2\Gamma)},
$$ 
where $\Delta_{00}$ is the zero-temperature gap of the same system
in absence of pair breaking. Note that the critical pair-breaking rate
for destruction of superconductivity is therefore
$\Gamma_{\rm max}=\Delta_{00}/2$.

Making use of the Sommerfeld-like expansion on the right-hand side of
Eq.~\eqref{eq:self-consistent}, after some algebra one finds that at
low temperatures
$$
\frac{\Delta(T)}{\Delta(0)}=1-\frac{\pi^2}{6}
\frac{\Gamma T^2}{(\Delta_{00}-\Gamma)^2(\Delta_{00}-2\Gamma)}.
$$ 
Note that, for finite $\Gamma$, the gap diminishes as a square of
temperature, very unlike the exponential behavior of the BCS
theory. This might be observable experimentally, but obviously very
high-precision data would be needed for that purpose.

{\it Superfluid fraction.} As shown in Ref.~\onlinecite{Herman17b},
the superfluid fraction of a Dynes superconductor at finite
temperature is given by
$$
\frac{n_s(T)}{n}=2\pi T\sum_{\omega_n>0}
\frac{\Delta^2}{\Omega_n^2(\Omega_n+\Gamma_s)}.
$$ 
Making use of the Sommerfeld-like expansion and noting that
$\Delta$ itself is temperature-dependent, after straightforward but
tedious calculations one can show that, similarly as the
superconducting gap, also the superfluid fraction diminishes as a
square of temperature, provided the pair-breaking rate is finite. The
temperature dependence of the superfluid fraction should be much
easier to test experimentally.

The low-temperature expansion of the superfluid fraction is in the
general case given by a cumbersome expression, but in the dirty limit
$\Gamma_s\gg\Delta_{00}$ it reduces to
\begin{equation}
\frac{n_s(T)}{n}=
\frac{\Delta(0)}{\Gamma_s}
\arctan\left(\frac{\Delta(0)}{\Gamma}\right)
-\frac{\pi^2}{6}\frac{\Gamma}{\Gamma_s}
\frac{\kappa T^2}{(\Delta_{00}-\Gamma)^2},
\label{eq:fraction}
\end{equation}
where the dimensionless coefficient $\kappa$ is a weak function of
$\Gamma/\Delta_{00}$ which varies between 3.57 and 4.1.  An explicit
formula for $\kappa$ is given in Appendix~B.

{\it Thermodynamic critical field.}  Applying the Sommerfeld-like
expansion to Eq.~\eqref{eq:free_diff}, we obtain
\begin{equation}
\delta{\cal F}=
-\frac{1}{2}N_0\left(\Delta_{00}-2\Gamma\right)^2
+\frac{\pi^2}{3}\frac{\Delta_{00}-2\Gamma}{\Delta_{00}-\Gamma}N_0T^2.
\label{eq:free_diff_expansion}
\end{equation}
The first term on the right-hand side is the condensation energy and
it changes smoothly from the well-known BCS value at $\Gamma=0$ to a
vanishing magnitude at the critical pair-breaking rate
$\Gamma=\Gamma_{\rm max}$.  The free energy difference $\delta{\cal
  F}$ per lattice site is related to the thermodynamic critical field
$H_c(T)$ by the well-known expression
\begin{equation}
\frac{\mu_0}{2}H_c^2=-\frac{\delta{\cal F}}{v_0},
\label{eq:hc}
\end{equation}
where $v_0$ is the unit-cell
volume. Equations~(\ref{eq:free_diff_expansion},\ref{eq:hc}) imply
that, similarly as the gap and the superfluid fraction, also the
thermodynamic critical field diminishes with the square of temperature
in the vicinity of absolute zero.\cite{note_hc}

\begin{figure*}[th!]
\includegraphics[width = 15 cm]{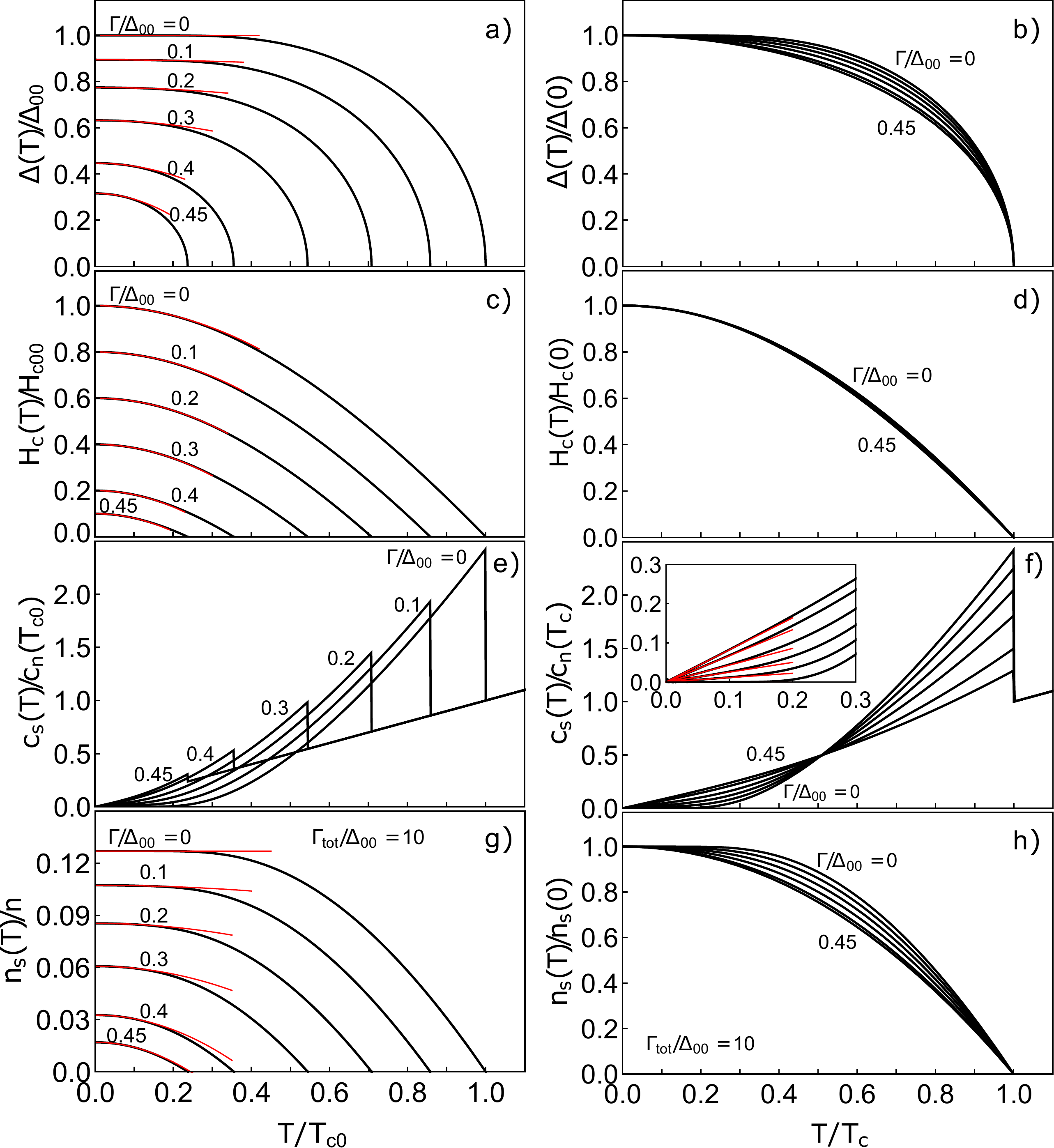}
\caption{Temperature dependence of thermodynamic quantities for the
  Dynes superconductors with several values of the pair-breaking rate
  $\Gamma$. The left column shows data in absolute (not
  $\Gamma$-dependent) units, whereas in the right column we use
  $\Gamma$-dependent units.  The low-temperature expansions are
  plotted as red curves. (a,b): the order parameter
  $\Delta(T)$. (c,d): critical field $H_c(T)$; $H_{c00}$ is the
  critical field of the clean system at $T=0$. (e,f): specific heat
  $c_S(T)$.  (g,h): superfluid fraction $n_s(T)$ for $\Gamma_{\rm
    tot}=10\Delta_{00}$.}
\label{fig:panel}
\end{figure*}

{\it Specific heat.} Combining the result
Eq.~\eqref{eq:free_diff_expansion} for the free energy difference with
the standard expression for the free energy of the normal metal,
$$
{\cal F}_N(T)={\cal F}_N(0)-\frac{\pi^2}{3}N_0T^2,
$$ 
we arrive at the following expression for the free energy of the
Dynes superconductor, valid in the limit of low temperatures:
$$
{\cal F}_S(T)={\cal F}_N(0)
-\frac{1}{2}N_0\left(\Delta_{00}-2\Gamma\right)^2
-\frac{\pi^2}{3}N(0)T^2,
$$ 
where $N(0)=\Gamma N_0/(\Delta_{00}-\Gamma)$ is the
zero-temperature limit of the density of states, evaluated at the
Fermi level.  The low-temperature specific heat $c_S=-T\partial^2{\cal
  F}_S/\partial T^2$ of a Dynes superconductor is therefore given by
$$
c_S=\frac{2\pi^2}{3}N(0)T.
$$ 
This means that the specific heat of the Dynes superconductor exhibits
the very same $T$-linear scaling as in the normal state, but with a
reduced density of states.  The full normal-state value is recovered
at the critical pair-breaking rate $\Gamma=\Gamma_{\rm max}$.

\subsection{Finite temperatures}
In what follows we study the full temperature dependence of the
thermodynamic quantities for several values of the pair-breaking
parameter $\Gamma$. The critical temperature of the clean system is
called $T_{c0}$, whereas $T_c$ denotes the critical temperature of the
same system with finite pair breaking.  The order parameter
$\Delta(T)$ is found by solving Eq.~\eqref{eq:self-consistent}, the
critical field $H_c(T)$ is determined making use of
Eqs.~(\ref{eq:free_diff},\ref{eq:hc}), and the electronic specific
heat in the superconducting state $c_S(T)$ is calculated using
Eq.~(\ref{eq:free_diff}) and
$$
c_S=\frac{2\pi^2}{3}N_0T-T\frac{\partial^2{\delta{\cal F}}}{\partial T^2}.
$$

Numerically obtained results for $\Delta(T)$, $H_c(T)$, and $c_S(T)$,
together with their low-temperature expansions, are shown in
Fig.~\ref{fig:panel}. It can be seen that the presence of a finite
pair-breaking rate $\Gamma$ leads, as expected, to reduced values of
$\Delta(T)$, $H_c(T)$, and $T_c$.

Note that when the critical field $H_c(T)$ is plotted in reduced
units, its overall temperature dependence changes only very slightly
in the whole range of allowed pair breaking rates between $0$ and
$\Gamma_{\rm max}$. Thus we can write
$$
\delta{\cal F}\approx
-\frac{1}{2}N_0\left(\Delta_{00}-2\Gamma\right)^2 f(\theta),
\qquad
\theta=\frac{T}{T_c}
$$ 
with a scaling function $f(\theta)$ which does not depend on
$\Gamma$ and is well approximated by a simple two-fluid form
$f(\theta)\approx (1-\theta^2)^2$. From the scaling formula for
$\delta{\cal F}$ it follows that the $c_S(T)/c_N(T_c)$ curves cross in
the vicinity of the reduced temperature $\theta_0$ given by the
equation $f''(\theta_0)=0$, as in fact observed in
Fig.~\ref{fig:panel}.

Finite pair breaking is most clearly visible in the electronic
specific heat $c_S(T)$: not only the low-temperature behavior exhibits
a qualitative change, but also the magnitude $\delta c(T_c)$ of the
specific heat jump at $T_c$ is strongly reduced by finite $\Gamma$:
$$
\frac{\delta c(T_c)}{c_N(T_c)}=
\frac{12\left[1
-\alpha\zeta\left(2,\frac{1}{2}+\alpha\right)\right]^2}
{\zeta\left(3,\frac{1}{2}+\alpha\right)},
\qquad
\alpha=\frac{\Gamma}{2\pi T_c},
$$
where $\zeta(s,x)=\sum_{n=0}^\infty 1/(n+x)^s$ is the Hurwitz zeta 
function. In the limit $\Gamma\rightarrow \Gamma_{\rm max}$, when
$\alpha\rightarrow\infty$, the specific heat jump vanishes as
$\delta c(T_c)/c_N(T_c)\approx 1/(6\alpha^2)$.

Also shown in Fig.~\ref{fig:panel} is the temperature dependence of
the superfluid fraction $n_s(T)$. Unlike other thermodynamic
quantities, $n_s(T)$ depends, in addition to $\Gamma$, also on the
value of the pair-conserving scattering rate $\Gamma_s$. We have
chosen $\Gamma_{\rm tot}=10\Delta_{00}$, corresponding to the
experimentally relevant dirty limit. Note that the low-temperature
power-law behavior of $n_s(T)$ should be clearly observable.

\subsection{Ginzburg-Landau region}
Since we know the temperature dependence of the free energy difference
$\delta{\cal F}$, the order parameter $\Delta$, and the superfluid
fraction $n_s/n$, close to $T_c$ we can construct the full
Ginzburg-Landau (GL) functional
\begin{equation}
\frac{\delta{\cal F}}{v_0}=
a |\psi|^2 +\frac{b}{2}|\psi|^4
+\frac{1}{2m^\ast}|\boldsymbol{\Pi}\psi|^2,
\label{eq:GL}
\end{equation}
where $m^\ast$ is the Cooper-pair mass, for which we take $m^\ast=2m$,
and $\boldsymbol{\Pi}=-i\hbar\nabla+2e{\bf A}$.

\subsubsection{Dirty limit}
In the dirty limit $\Gamma_s\gg\Delta_{00}$ we find that
Eq.~\eqref{eq:GL} obtains, if the GL wavefunction is taken as
$$
\psi=\sqrt{\tfrac{1}{4\pi}\zeta(2,\tfrac{1}{2}+\alpha)}
\frac{\Delta}{\sqrt{T_c\Gamma_s}}\sqrt{n}.
$$
The GL coefficients are given by the expressions
\begin{eqnarray*}
a&=&\frac{3\pi\left[1-\alpha\zeta(2,\tfrac{1}{2}+\alpha)\right]}
{\zeta(2,\tfrac{1}{2}+\alpha)}
\frac{(T-T_c)\Gamma_s}{\varepsilon_F},
\\
b&=&\frac{3\zeta(3,\tfrac{1}{2}+\alpha)}
{2\zeta^2(2,\tfrac{1}{2}+\alpha)}
\frac{\Gamma_s^2}{\varepsilon_F n},
\end{eqnarray*}
where $\varepsilon_F$ is the Fermi energy. From here it follows that
the GL coherence length $\xi$ and penetration depth $\lambda$ are
\begin{eqnarray*}
\frac{1}{\xi^2}&=&
\frac{24\left[1-\alpha\zeta(2,\tfrac{1}{2}+\alpha)\right]}
{\pi\zeta(2,\tfrac{1}{2}+\alpha)}
\frac{\Gamma_s}{\Delta_{00}}
\frac{T_c}{\Delta_{00}}
\frac{1-\theta}{\xi_0^2},
\\
\frac{1}{\lambda^2}&=&
\frac{4\pi\left[1-\alpha\zeta(2,\tfrac{1}{2}+\alpha)\right]
\zeta(2,\tfrac{1}{2}+\alpha)}
{\zeta(3,\tfrac{1}{2}+\alpha)}
\frac{\Delta_{00}}{\Gamma_s}
\frac{T_c}{\Delta_{00}}
\frac{1-\theta}{\lambda_0^2},
\end{eqnarray*}
where $\xi_0=\hbar v_F/(\pi\Delta_{00})$ and $\lambda_0^{-2}=\mu_0
ne^2/m$ are the coherence length and penetration depth of a system in
absence of all kinds of impurities at $T=0$. Note that, as is well
known, pair-conserving disorder, described by the factor
$\Gamma_s/\Delta_{00}$, renormalizes $\xi$ and $\lambda$ in opposite
ways.

In absence of pair breaking, i.e. for $\alpha=0$, the values of $\xi$
and $\lambda$ can be easily shown to coincide with the textbook
results for the dirty limit, see e.g. Ref.~\onlinecite{Tinkham04}.
One just has to take into account that the mean free path is given by
$\ell=\hbar v_F/(2\Gamma_s)$.

In the opposite limit $\Gamma\rightarrow\Gamma_{\rm max}$, when $T_c$
vanishes and $\alpha\rightarrow\infty$, the expressions for $\xi$ and
$\lambda$ reduce to
\begin{eqnarray*}
\frac{1}{\xi^2}&=&8\frac{\Gamma_s}{\Delta_{00}}
\left(\frac{T_c}{\Delta_{00}}\right)^2
\frac{1-\theta}{\xi_0^2},
\\
\frac{1}{\lambda^2}&=&\frac{8\pi^2}{3}
\frac{\Delta_{00}}{\Gamma_s}
\left(\frac{T_c}{\Delta_{00}}\right)^2
\frac{1-\theta}{\lambda_0^2}.
\end{eqnarray*}
This means that both, $\xi$ and $\lambda$, increase by the same large
factor $\Delta_{00}/T_c$ when the pair-breaking rate $\Gamma$
increases. Also the magnitude of the thermodynamic critical field,
$$
\mu_0 H_c=\sqrt{\frac{8}{3}}\left(\frac{T_c}{\Delta_{00}}\right)^2
\frac{\Phi_0}{\lambda_0\xi_0}(1-\theta),
$$ 
is suppressed by the large factor $(\Delta_{00}/T_c)^2$.

On the other hand, the GL parameter $\kappa=\lambda/\xi\sim
(\Gamma_s/\Delta_{00})\times \lambda_0/\xi_0$ changes only little from
$\kappa=0.72\lambda_0/\ell$ in absence of pair breaking to
$\kappa=0.87\lambda_0/\ell$ at $\Gamma\rightarrow\Gamma_{\rm max}$.
Therefore the difference between the upper and lower critical fields,
$H_{c2}/H_c\approx H_c/H_{c1}\approx \sqrt{2}\kappa$, is controlled
essentially by the factor $\Gamma_s/\Delta_{00}$.

\subsubsection{Clean limit}
For the sake of completeness, let us also study superconductors with a
vanishing pair-conserving scattering rate $\Gamma_s=0$ and a finite
pair breaking rate $\Gamma$. Such superconductors will be called clean
in what follows.  In this case the GL wavefunction has to be taken as
$$
\psi=\sqrt{\tfrac{1}{2}\zeta(3,\tfrac{1}{2}+\alpha)}
\frac{\Delta}{2\pi T_c}\sqrt{n}
$$
and the GL coefficients are given by the expressions
\begin{eqnarray*}
a&=&\frac{6\pi^2\left[1-\alpha\zeta(2,\tfrac{1}{2}+\alpha)\right]}
{\zeta(3,\tfrac{1}{2}+\alpha)}
\frac{(T-T_c)T_c}{\varepsilon_F},
\\
b&=&\frac{6\pi^2}{\zeta(3,\tfrac{1}{2}+\alpha)}
\frac{T_c^2}{\varepsilon_F n}.
\end{eqnarray*}
From here it follows that $\xi$ and $\lambda$ are given by
\begin{eqnarray*}
\frac{1}{\xi^2}&=&
\frac{48\left[1-\alpha\zeta(2,\tfrac{1}{2}+\alpha)\right]}
{\zeta(3,\tfrac{1}{2}+\alpha)}
\left(\frac{T_c}{\Delta_{00}}\right)^2
\frac{1-\theta}{\xi_0^2},
\\
\frac{1}{\lambda^2}&=&
2\left[1-\alpha\zeta(2,\tfrac{1}{2}+\alpha)\right]
\frac{1-\theta}{\lambda_0^2}.
\end{eqnarray*}
One checks readily that, in absence of pair breaking, i.e. for
$\alpha=0$, these formulas reproduce the textbook results for clean
superconductors, see e.g. Ref.~\onlinecite{Tinkham04}. 

On the other hand, in the opposite limit of strong pair breaking,
$\Gamma\rightarrow\Gamma_{\rm max}$, we find
\begin{eqnarray*}
\frac{1}{\xi^2}&=&
8\left(\frac{T_c}{\Delta_{00}}\right)^2
\frac{1-\theta}{\xi_0^2},
\\
\frac{1}{\lambda^2}&=&
\frac{8\pi^2}{3}
\left(\frac{T_c}{\Delta_{00}}\right)^2
\frac{1-\theta}{\lambda_0^2}.
\end{eqnarray*}
Note that, similarly as in the dirty limit, pair-breaking disorder
renormalizes both, $\xi$ and $\lambda$, by essentially the same
factor. This is quite different from the effect of pair-conserving
disorder.

\begin{figure}[t!]
\includegraphics[width = 7 cm]{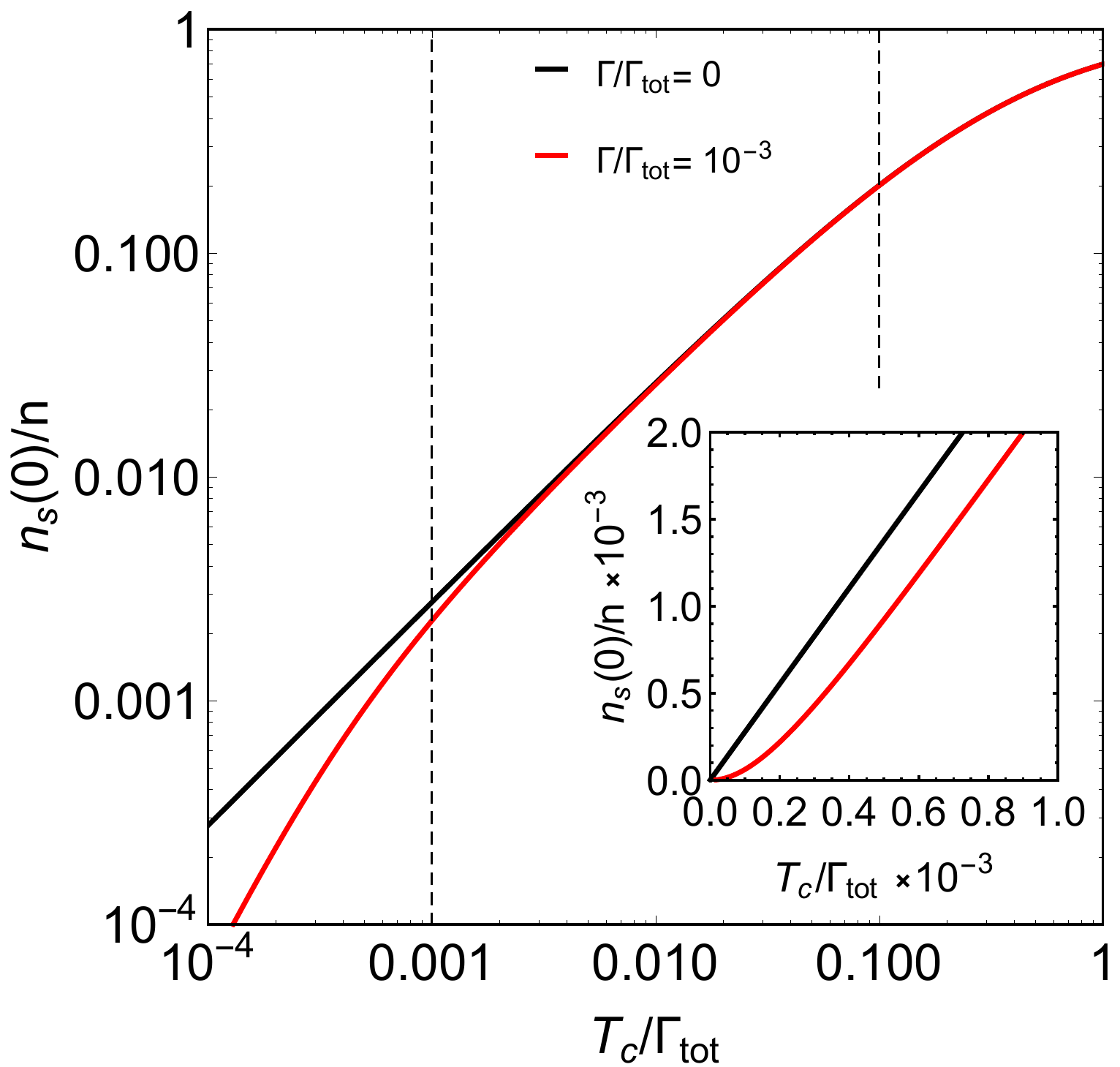}
\caption{Superfluid fraction $n_s(0)/n$ for a series of Dynes
  superconductors with varying $T_c$. The total scattering rate
  $\Gamma_{\rm tot}=\Gamma_s+\Gamma$ is kept fixed and two
  pair-breaking scattering rates $\Gamma\ll \Gamma_{\rm tot}$ are
  considered. Three regimes are visible, separated by the vertical
  dashed lines: clean regime with $\Gamma_{\rm tot}\lesssim T_c$, the
  Homes regime $\Gamma\lesssim T_c\lesssim \Gamma_{\rm tot}$, and the
  novel pair-breaking regime $T_c\lesssim \Gamma$.  }
\label{fig:homes}
\end{figure}

\section{Modified Homes' law}
Superconductivity often occurs in the vicinity of quantum critical
points. The observed dome-like shapes of the phase diagrams are then
usually explained by the assumption that the pairing strength
decreases with distance from the critical point. The control parameter
measuring this distance may be pressure, or - as in the case of the
cuprates - doping. Recently, Bo\v{z}ovi\'{c} et al. have studied the
scaling of the low-temperature value of the superfluid density
$n_s(0)$ with $T_c$ in overdoped cuprates.\cite{Bozovic16} They have
found that when the doping was increased, i.e. when the coupling
constant and $T_c$ decreased, $n_s(0)$ went down, initially following
the Homes law\cite{Dordevic13} $n_s(0)/n\propto T_c/\Gamma_{\rm tot}$,
but when $T_c$ dropped below $\sim 10$~K, a different scaling
$n_s(0)/n \propto (T_c/\Gamma_{\rm tot})^2$ was
observed.\cite{Bozovic16}

At first sight, the vanishing of $n_s(0)$ looks mysterious, since in a
clean system the superfluid density should be given by the density of
electrons $n$, which obviously does not vanish in the samples studied
by Bo\v{z}ovi\'{c} et al. In a clean system, one should therefore
expect that $n_s(0)=n$ holds for all samples with a finite critical
temperature $T_c$, and $n_s(0)$ should jump discontinuosly (!) to zero
when $T_c=0$. However, in any real sample, disorder is present.  This
implies that, when the critical temperature enters the range
$T_c\lesssim \Gamma_{\rm tot}=\Gamma_s+\Gamma$, the sample is in the
dirty limit and the Homes scaling
\begin{equation}
\frac{n_s(0)}{n}\approx \frac{\pi}{2}\frac{\Delta(0)}{\Gamma_{\rm tot}}
\label{eq:homes}
\end{equation} 
or $n_s(0)\propto T_c$ should apply, resulting in a smooth
development of $n_s(0)$ across the superconductor to normal metal
transition.
 
In Fig.~\ref{fig:homes} the superfluid fraction $n_s(0)/n$ for a
series of Dynes superconductors with varying coupling strength $g$,
but fixed scattering rates $\Gamma$ and $\Gamma_{\rm tot}$, is plotted
as a function of their transition temperature $T_c$, making use of the
explicit formula for $n_s(0)/n$ given in Ref.~\onlinecite{Herman17b}.
In order to emphasize the power-law behavior, the log-log plot is
used; the curve for $\Gamma=0$ coincides with the result of
Ref.~\onlinecite{Kogan13}.  As expected, in the intermediate range of
temperatures $\Gamma\lesssim T_c\lesssim \Gamma_{\rm tot}$ the Homes
scaling Eq.~\eqref{eq:homes} does apply. However, it turns out - see
also Eq.~\eqref{eq:fraction} - that very close to the superconductor -
normal metal transition, there exists yet another regime, $T_c\lesssim
\Gamma$, where the Homes scaling is replaced by
\begin{equation}
\frac{n_s(0)}{n}\approx \frac{\Delta^2(0)}{\Gamma_{\rm tot}\Gamma}
\label{eq:homes_modified}
\end{equation}
and therefore $n_s(0)\propto T_c^2$, exactly as observed in overdoped
cuprates by Bo\v{z}ovi\'{c} et al. A similar result has been obtained
previously,\cite{Kogan13b} but in that work only weak impurity
scattering was considered and the Born approximation was used.

One should note that the result Eq.~\eqref{eq:homes_modified}
obviously can not be directly applied to the cuprates with d-wave
pairing symmetry. On the other hand, since any real samples should
exhibit a finite (although probably small) value of the pair-breaking
rate $\Gamma$, the anomalous scaling $n_s(0)\propto T_c^2$ in the
immediate vicinity of the superconductor - normal metal transition
should be generic.

\section{Conclusions}
In this paper we have dealt with the recently introduced Dynes
superconductors, i.e.  with superconductors with simultaneously
present pair-conserving and pair-breaking scattering processes with
rates $\Gamma_s$ and $\Gamma$, described by the Green's function given
by Eqs.~(\ref{eq:dynes_green},\ref{eq:epsilon}).

First we have demonstrated, following the classic work by
Jani\v{s},\cite{Janis89} that the theory of the Dynes superconductors
is thermodynamically consistent, since the CPA equations on which it
is based can be derived from a free-energy functional of the
self-energy. 

We have derived a convenient expression for the free energy difference
between the superconducting and the normal state,
Eq.~\eqref{eq:free_diff}, and we have shown that it respects
Anderson's theorem. Making use of Eq.~\eqref{eq:free_diff}, we have
calculated the electronic specific heat and the thermodynamic critical
field of the Dynes superconductors.

Our main result is that the gap parameter, specific heat, critical
field, and penetration depth of the Dynes superconductors exhibit
power-law scaling with temperature in the low-temperature limit. These
results follow from the gapless nature of the Dynes superconductors
and they should be readily falsifiable by experiments.

Close to $T_c$, we have constructed the Ginzburg-Landau functional for
the Dynes superconductors. Our most interesting observation is that
(at least in this region of temperatures) the two types of scattering
processes influence the upper critical field $H_{c2}$ in exactly
opposite ways: In agreement with textbook results, the pair-conserving
processes lead to an {\it increase} of $H_{c2}$, essentially due to a
suppressed sensibility of the electrons to magnetic field. On the
other hand, pair-breaking processes lead to a {\it suppression} not
only of the thermodynamic critical field $H_c$ and of the lower
critical field $H_{c1}$, but also of $H_{c2}$.

We have also shown that, in the immediate vicinity of a coupling
constant-controlled superconductor to normal metal transition, the
Homes law Eq.~\eqref{eq:homes} is generically replaced by the
pair-breaking dominated scaling law Eq.~\eqref{eq:homes_modified}.
Although a similar result has in fact been found recently in overdoped
cuprates,\cite{Bozovic16} our theory does not apply directly to that
experiment, since the cuprates are d-wave superconductors, whereas our
theory considers s-wave pairing symmetry.  Nevertheless, we speculate
that the distinction between pair-conserving (small-angle) and
pair-breaking (large-angle) scattering might play a role in the
experiment of Bo\v{z}ovi\'{c} et al.  At least the starting point
seems to work: small-angle scattering is known to dominate over the
large-angle scattering in the cuprates.\cite{Herman17,Reber12,Hong14}
However, a serious analysis of the results of
Ref.~\onlinecite{Bozovic16} is beyond the scope of this work.

\begin{acknowledgments}
This work was supported by the Slovak Research and Development Agency
under contracts No.~APVV-0605-14 and No.~APVV-15-0496, and by the
Agency VEGA under contract No.~1/0904/15. F.H. is grateful for the 
financial support to the Swiss National Science Foundation.
\end{acknowledgments}

\appendix
\section{Calculation of $\delta{\cal F}$}
In this Appendix we will show that the contributions of the second and
of the third terms in Eq.~\eqref{eq:free} to $\delta{\cal F}$ of a
Dynes superconductor can be neglected.  To this end, let us first note
that the local Green's function of a Dynes superconductor is given
by\cite{Herman16}
$$
\hat{\cal G}_n^{-1}=\frac{1}{\pi N_0}
\frac{i\omega_n\tau_0-\Delta_n\tau_1}{\sqrt{\omega_n^2+\Delta_n^2}},
$$ 
and the corresponding normal-state local Green's function obtains
from the same expression by setting $\Delta_n=0$.

Let us start by considering the contribution of the second term in
Eq.~\eqref{eq:free} to the free energy.  Making use of the identity
\begin{equation}
{\rm Tr}\ln(A\tau_0+B\tau_1+C\tau_3)=\ln(A^2-B^2-C^2)
\label{eq:identity}
\end{equation}
one checks readily that this contribution is the same in both, the
normal and the superconducting states. Therefore the second term in
Eq.~\eqref{eq:free} does not contribute to $\delta{\cal F}$.

\begin{widetext}
Thus we are left with the contribution of the third term to
$\delta{\cal F}$. Making use of Eq.~\eqref{eq:identity}, this can be
written as
\begin{eqnarray}
-T\sum_n\left\langle \ln\left[1+
2\left(\frac{|\omega_n|}{\sqrt{\omega_n^2+\Delta_n^2}}-1\right)
\frac{(1-\pi N_0\Gamma_s)(i\lambda-\pi N_0\Gamma)}
{\mu^2+(1-\pi N_0\Gamma_{\rm tot}+i\lambda)^2}
\right]\right\rangle,
\label{eq:3rdterm}
\end{eqnarray}
where we have introduced dimensionless pair-conserving and
pair-breaking fields $\mu=\pi N_0 U$ and $\lambda=\pi N_0 V$,
respectively. We remind that the angular brackets denote averaging
with respect to the random fields $U$ and $V$.  In deriving
Eq.~\eqref{eq:3rdterm}, we have assumed that the pair-breaking
distribution function $P_m(V)$ is even. Noting that the expression in
round brackets under the logarithm is proportional to $\Delta^2$,
Eq.~\eqref{eq:3rdterm} can be written, to order $\Delta^2$, in the
form
\begin{eqnarray}
-2T\sum_n
\left(\frac{|\omega_n|}{\sqrt{\omega_n^2+\Delta_n^2}}-1\right)
\left\langle \frac{(1-\pi N_0\Gamma_s)(i\lambda-\pi N_0\Gamma)}
{\mu^2+(1-\pi N_0\Gamma_{\rm tot}+i\lambda)^2}\right\rangle.
\label{eq:3rdterm2}
\end{eqnarray}
\end{widetext}
A simple integration in complex plane shows that, for a Lorentzian
distribution of the pair-breaking field $V$ with width $\Gamma$, and
for not too strong disorder, $\pi N_0\Gamma_{\rm tot}<1$, the average
in Eq.~\eqref{eq:3rdterm2} vanishes. This means that the expression
Eq.~\eqref{eq:3rdterm} is proportional at least to $\Delta^4$ and
therefore clearly negligible with respect to the result
Eq.~\eqref{eq:free_diff}, as claimed in the main text.

\section{Sommerfeld-like expansion}
In this Appendix we will derive the Sommerfeld-like expansion.  Let us
start by observing that we can always split the frequency integral of
any function $F(\omega)$ into the following sum of  integrals:
$$
\int_0 d\omega F(\omega)=\sum_{n=0}
\int_{2\pi Tn}^{2\pi T(n+1)} d\omega F(\omega).
$$ 
If the temperature $T$ is small, then the interval $\langle 2\pi T
n, 2\pi T (n+1)\rangle$ is short and we can calculate the
corresponding integral by Taylor expansion of $F(\omega)$ around the
center of the interval, which happens to coincide with the fermionic
Matsubara frequency $\omega_n=(2n+1)\pi T$. This way we obtain, to
order $T^4$,
$$
\int_{2\pi Tn}^{2\pi T(n+1)} d\omega F(\omega)=
2\pi T\left[F(\omega_n)+\frac{\pi^2}{6} T^2 F''(\omega_n)\right].
$$
Note that even powers of $T$ do not enter this expansion.
Summing these results and restoring the upper limit of integration we
obtain
\begin{equation}
\int_0^{\Omega} d\omega F(\omega)=
2\pi T\sum_{\omega_n>0}^{\Omega}
\left[F(\omega_n)+\frac{\pi^2}{6} T^2 F''(\omega_n)\right].
\label{eq:aux_b}
\end{equation}
Since the sum on the right-hand side contains $\propto T^{-1}$ terms,
Eq.~\eqref{eq:aux_b} represents the integral to order $T^3$.

At this point it is sufficient to realize that, to order $T$, the
same argument leads to the result
\begin{equation}
\int_0^{\Omega} d\omega F''(\omega)=
2\pi T\sum_{\omega_n>0}^{\Omega} F''(\omega_n).
\label{eq:aux_b2}
\end{equation}
If we make use of the result Eq.~\eqref{eq:aux_b2} on the right-hand
side of Eq.~\eqref{eq:aux_b}, after a trivial manipulation we finally
arrive at the Sommerfeld-like expansion, valid to order $T^3$,
$$
2\pi T\sum_{\omega_n>0}^{\Omega} F(\omega_n)=
\int_0^{\Omega} d\omega F(\omega)
+\frac{\pi^2}{6}T^2\left[F'(0)-F'(\Omega)\right].
$$

Note that usually $F'(\Omega)$ can be neglected, and therefore a
finite $T^2$ correction is present only if $F'(0)$ does not vanish.
Since $F(\omega)$ is typically an even function, this means that a
finite $T^2$ correction results only if $F(\omega)$ is non-analytic at
$\omega=0$. This is indeed the case for the Dynes superconductors with
a finite value of the pair-breaking scattering rate $\Gamma$. For the
sake of completeness let us note that for some quantities, $F'(0)$ may
be finite also in the clean BCS case. For example, this is the case
for $\delta{\cal F}$, see Eq.~\eqref{eq:free_diff}.

The low-temperature expansions presented in Section~III.A are obtained
by repeated use of the Sommerfeld-like expansion. For the sake of
completeness let us mention that the dimensionless coefficient
$\kappa$ which appears in the expansion of the superfluid fraction is
given by the expression
\begin{eqnarray*}
\kappa=
\frac{\Delta_{00}}{\Delta(0)}\arctan\left(\frac{\Delta(0)}{\Gamma}\right)
+\frac{\Delta_{00}}{\Delta_{00}-\Gamma}
+\left(\frac{\Delta(0)}{\Delta_{00}-\Gamma}\right)^2.
\end{eqnarray*}


\begin{thebibliography}{99}
\bibitem{note_gap_equation}At this point it is convenient to consider
  a complex pairing field $\Delta$, and to treat $\hat{V}_{12}=\Delta$
  and $\hat{V}_{21}=\Delta^\ast$ as
  independent. Equation~\eqref{eq:self-consistent} obtains by
  minimization with respect to $\Delta^\ast$. In the rest of this
  paper we work in a gauge where the pairing field is real.  The
  frequency cutoff $\Omega_{\rm max}$ in
  Eq.~\eqref{eq:self-consistent} has been introduced by hand.
\bibitem{note_1stterm}We have made use of the identity
  Eq.~\eqref{eq:identity}.
\bibitem{note_hc}Note that $H_c(T)$ exhibits the same low-temperature
  behavior in both, the Dynes and the BCS superconductors, see
  Appendix~B.
\end{thebibliography}
\end{document}